\documentclass{PoS}
\usepackage{graphicx}
\usepackage{wrapfig}
\def\gagg{g_{a\gamma}}
\def\iso#1#2{\leavevmode\hbox{$^{#1}$#2}}
\def\etal{{\it et al.}}

\title{Tokyo axion helioscope experiment (20'+5')}

\ShortTitle{Tokyo axion helioscope experiment (20'+5')}

\author{\speaker{Yoshizumi~INOUE}$^{ad}$,
  Yuki~AKIMOTO$^b$,
  Ryosuke~OHTA$^b$,
  Tetsuya~MIZUMOTO$^b$,
  Akira~YAMAMOTO$^c$,
  and
  Makoto~MINOWA$^{bd}$
  \\
  \llap{$^a$}International Center for Elementary Particle Physics (ICEPP),
  University of Tokyo\\
  7-3-1~Hongo, Bunkyo-ku, Tokyo~113-0033, Japan\\
  \llap{$^b$}Department of Physics, School of Science, University of Tokyo\\
  7-3-1~Hongo, Bunkyo-ku, Tokyo~113-0033, Japan\\
  \llap{$^c$}High Energy Accelerator Research Organization (KEK)\\
  1-1~Oho, Tsukuba, Ibaraki~305-0801, Japan\\
  \llap{$^d$}Research Center for the Early Universe (RESCEU),
  School of Science, University of Tokyo\\
  7-3-1~Hongo, Bunkyo-ku, Tokyo~113-0033, Japan\\
  E-mail: \email{berota@icepp.s.u-tokyo.ac.jp}}

\abstract{
  A search for solar axions has been performed using an axion helioscope
  which is equipped with a $2.3\,{\rm m}\times4\rm\,T$ superconducting
  magnet, a gas container to hold dispersion-matching gas, PIN-photodiode
  X-ray detectors, and a telescope mount mechanism to track the sun.
  In the past measurements, axion mass up to 0.27\,eV have been scanned
  with this helioscope.
  It has been upgraded to handle dispersion-matching gas ($^4$He) of
  higher density to explore higher mass region.
  From December 2007 through April 2008, the axion mass region
  $0.84<m_a<1.00\rm\,eV$ was newly explored, where the axions in the
  ``photon-coupling vs. mass'' parameter region ($\gagg$-$m_a$) of the
  preferred axion models were newly searched by this experiment.
  From the absence of any evidence, a limit on axion-photon coupling
  constant was set to be
  $\gagg<\hbox{5.6--13.4}\times10^{-10}{\rm GeV}^{-1}$ at 95\%
  confidence level in the above mass region.
}

\FullConference{Identification of dark matter 2008\\
		 August 18-22, 2008\\
		 Stockholm, Sweden}

\begin{document}

\section{Introduction}
\begin{wrapfigure}{r}{0.45\textwidth}
  \vbox{\hrule height0pt
    \vskip -2pc
    \hbox{\includegraphics[width=0.4\textwidth]{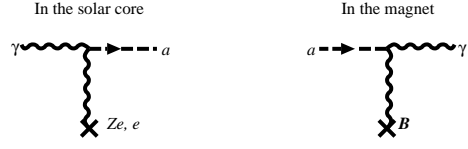}}}
  \caption{The solar axions produced via the Primakoff process in the
    solar core are, then, converted into X-rays via the reverse process
    in the magnet.}
  \label{fig:principle}
\end{wrapfigure}  
The axion is a Nambu--Goldstone boson of a spontaneously broken U(1)
symmetry, the Peccei--Quinn symmetry, which was introduced to solve
the strong CP problem in quantum chromodynamics (QCD)
\cite{axion-bible1}.
If its mass $m_a$ is at around a few electronvolts, the sun can be
a powerful source of axions and the so-called `axion helioscope'
technique may enable us to detect such axions
\cite{sikivie1983,bibber1989}.

The principle of the axion helioscope is illustrated in
Fig.~\ref{fig:principle}.
Axions are expected to be produced in the solar core through the
Primakoff process.
The average energy of the solar axions is 4.2\,keV and their
differential flux expected at the Earth is approximated by
\cite{bahcall2004,raffelt2005}
\begin{eqnarray}
  {\rm d}\Phi_a/{\rm d}E&=&6.020\times10^{10}[\mathrm{cm^{-2}s^{-1}keV^{-1}}]
  \nonumber\\
  &&{}\times\left(\gagg\over10^{-10}\mathrm{GeV}^{-1}\right)^2
  \left( {E\over1\rm\,keV} \right)^{2.481}
  \exp \left( -{E\over1.205\rm\,keV} \right),
  \label{eq:aflux}
\end{eqnarray}
where $E$ is the energy of the axions and $\gagg$ is the axion-photon
coupling constant.
Then, they would be converted into X-ray photons through the inverse
process in a strong magnetic field at a laboratory.
The conversion rate is given by
\begin{equation}
P_{a\to\gamma} = {\gagg^2\over4} 
  \exp\left[-\int_0^L\!\!\mathrm{d}z\,\Gamma \right]
  \times
  \left| \int_0^L\!\!{\rm d}z\,B_\bot\exp
  \left[i \int_0^z\!\!\mathrm{d}z^\prime
    \left(q - {i\Gamma\over2}\right) \right]\right|^2,
  \label{eq:prob}
\end{equation}
where $z$ and $z^\prime$ are the coordinate along the incident solar
axion, $B_\bot$ is the strength of the transverse magnetic field,
$L$ is the length of the field along $z$-axis, $\Gamma$ is the X-ray
absorption coefficient of the filling medium,
$q=(m_\gamma^2-m_a^2)/2E$ is the momentum transfer by the virtual
photon, and $m_\gamma$ is the effective mass of the photon in medium.
In light gas, such as hydrogen or helium, $m_\gamma$ is well
approximated by
\begin{equation}
  m_\gamma=\sqrt{4\pi\alpha N_e/m_e},
\end{equation}
where $\alpha$ is the fine structure constant, $m_e$ is the electron
mass, and $N_e$ is the number density of electrons.
In vacuum, the axion helioscope is not sensitive to massive axions due
to a loss of coherence by non-zero $q$.
However, coherence can be restored if one can adjust $m_\gamma$ to $m_a$.

We adopted cold \iso{4}{He} gas as a dispersion-matching medium.
The gas was kept at almost the same temperature as the magnet,
$T\lesssim6\rm\,K$, which is just above the critical temperature of
\iso{4}{He}, $T_c=5.2\rm\,K$.
It is worth noting that axions as heavy as a few electronvolts can be
reached with helium gas of only about one atmosphere and \iso{4}{He}
will not liquify at any pressure at this temperature.

In 1997, the first phase measurement \cite{sumico1997} was performed
without the gas container.
In this measurement, the sensitive mass region was limited to
$m_a<0.03\rm\,eV$ since the conversion region was vacuum.
In 2000, the second phase measurement \cite{sumico2000} was performed
to search for sub-electronvolt axions.
This experiment, together with the first phase measurement
\cite{sumico1997}, yielded an upper limit of
$\gagg<\hbox{6.0--10.5}\times10^{-10}\rm GeV^{-1}$ (95\% CL) for
$m_a<0.27\rm\,eV$.

Here, we will present the result of the third phase measurement in
which the mass region around 1\,eV was scanned using the upgraded
apparatus to withstand higher pressure gas.

\section{Experimental apparatus}
\begin{wrapfigure}{r}{0.45\textwidth}
  \vbox{\hrule height 0pt
    \vskip -2pc
    \hbox{\includegraphics[width=0.45\textwidth]{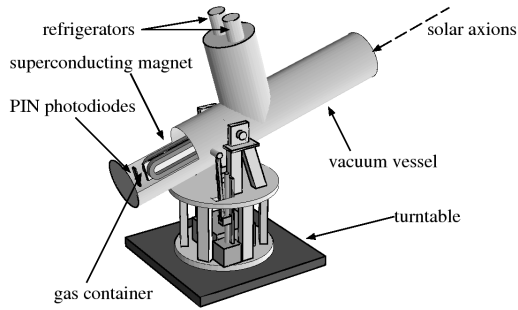}}%
    \hrule height 0pt}
  \caption{The schematic view of the axion helioscope called
    the Sumico V detector.}
  \label{fig:sumico}
\end{wrapfigure}
The axion helioscope consists of a superconducting magnet, X-ray
detectors, a gas container, and an altazimuth mounting to track the sun.
The schematic figure is shown in Fig.~\ref{fig:sumico}.

The superconducting magnet \cite{sato1997} consists of two 2.3-m long
race-track shaped coils running parallel with a 20-mm wide gap between
them.
The magnetic field in the gap is 4\,T perpendicular to the helioscope
axis.
The coils are kept at 5--6\,K during operation.
The magnet is made cryogen-free by two Gifford-McMahon refrigerators
cooling directly by conduction, and is equipped with a persistent
current switch.

The container to hold dispersion-matching gas is inserted in the gap
of the magnet.
Its body is made of four stainless-steel square pipes welded side by
side to each other, and is wrapped with 5N high purity aluminium sheet
to achieve high uniformity of temperature.
The uniformity of the temperature guarantees the homogeneous density
along the length of the container.
The detector side of the container is ended with an X-ray window which
is transparent to X-ray above 2\,keV and can withstand up to 0.3\,MPa.
The container is fixed to the magnet at this side through a
temperature-stablilized thermal linkage.
The solar end at the opposite side is blind-ended and is suspended by
three Kevlar cords, so that thermal flow through this end is highly
suppressed.

\hyphenation{ho-ri-ba-stec}
To have automatic sequential pressure settings, a gas handling system
is built with HORIBASTEC Piezo valves and a YOKOGAWA precision
pressure gauge.
For emergency exhaust of the gas in case of rapid temperature increase
due to a magnet quenching, a cryogenic rupture disk, which is designed
to break at 0.248 MPa, is also introduced into the gas handling system
to avoid destruction of the X-ray window by the over pressure.

Sixteen PIN photodiodes, Hamamatsu Photonics S3590-06-SPL, are used as
the X-ray detectors \cite{naniwaPIN}.
In the present measurement, however, twelve of them are used for the
analysis because four went defective through thermal stresses.
The chip size of a photodiode is $11\times11\times0.5\rm\,mm^3$, and
the effective area is larger than $9\times9\,\mathrm{mm^2}$.
It has an inactive surface layer of $0.35\,\mu\mathrm{m}$
\cite{akimotoPIN}.
The output from each photodiode is fed to a charge sensitive
preamplifier and waveforms of the preamplifier outputs are digitized
using FADCs.
We applied off-line pulse shaping to the recorded waveforms as
described in Ref.~\cite{sumico2000}.
Each photodiode was calibrated by 5.9-keV Mn X-rays from a \iso{55}{Fe}
source installed in front of them.
The source is manipulated from the outside and is completely retracted
behind the shield during the observations.

The entire axion detector is constructed in a vacuum vessel and the
vessel is mounted on a computer-controlled altazimuth mount.
Its trackable altitude ranges from $-28^\circ$ to $+28^\circ$ and its
trackable azimuthal range is almost $360^\circ$.
However, in the present measurement, the azimuthal range without a
human intervention is restricted to about 60$^\circ$ because a cable
handling system for its unmanned operation is not completed yet.
This range corresponds to an exposure time of about a quarter of a
day in observing the sun.
During the rest three quarters of a day, background was measured.

\section{Measurement and Analysis}
From December 21 2007 through April 21 2008, a new measurement was
performed for 34 gas-density settings with about three days of running
time per setting.
The scanned mass range was 0.84--1 eV.
Since we had not completed the gas relief system, the highest density
was determined so that the gas pressure would not exceed the breakage
pressure of the rupture disk even during a magnet quenching.

Energy spectra for the solar observation and the background are
obtained for each density settings based on the measured direction of
the helioscope.
Event reduction process was applied in the same way as described in
Ref.~\cite{sumico2000}.
We searched for expected axion signals which scale with $\gagg^4$ for
various $m_a$ in these spectra by applying a series of least $\chi^2$
fittings assuming various $m_a$ values.
Data from the 34 different gas density settings were combined by using
the summed $\chi^2$ of the 34.
The energy region of 4--20\,keV was used for fitting.
As a result, no significant excess was seen for any $m_a$, and thus
an upper limit on $\gagg$ at 95\% confidence level was given.

\begin{figure}
  \includegraphics[scale=0.85]{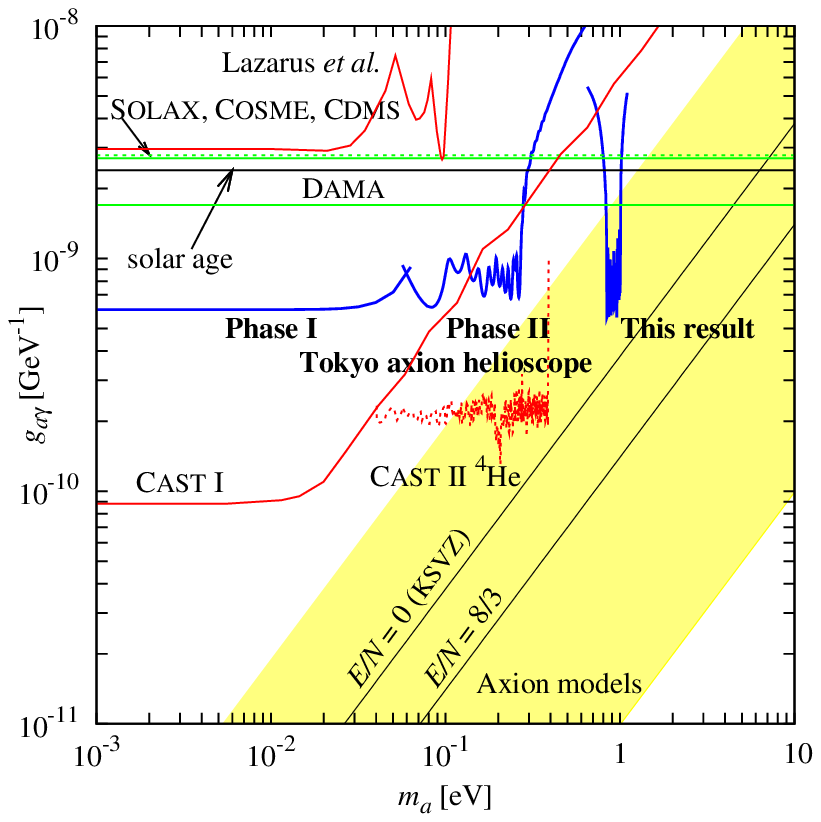}
  \hfill
  \includegraphics[scale=0.85]{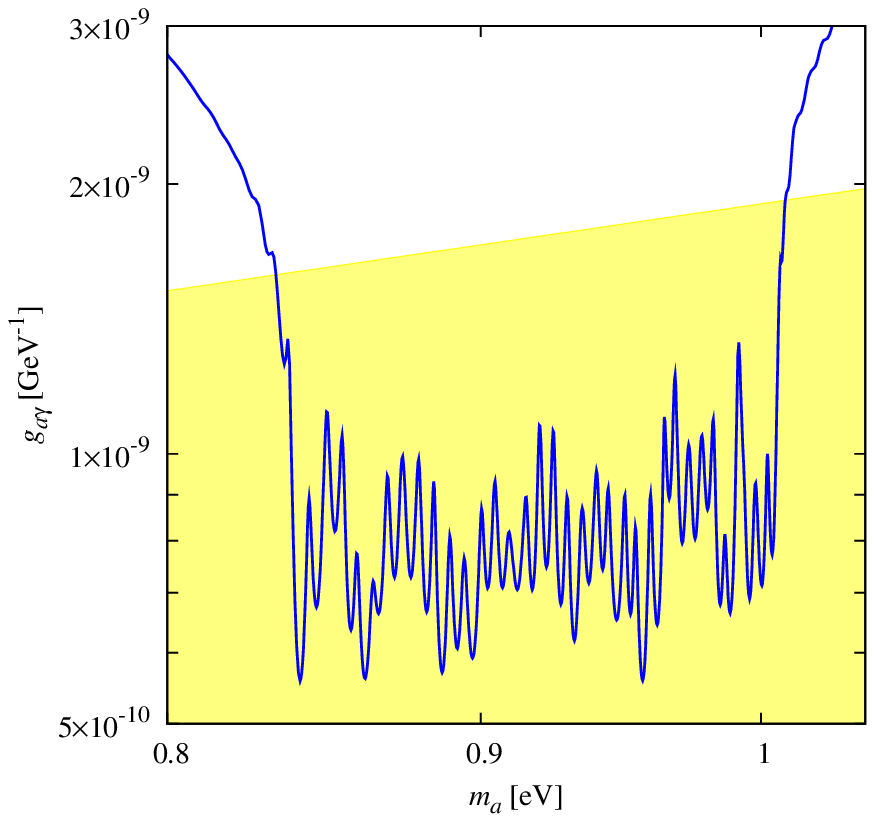}
  \caption{
    The left figure show the exclusion plot for $\gagg$ as a function
    of $m_a$.
    The shaded area corresponds to the preferred axion models
    \cite{GUT_axion}.
    Two slopes in the area correspond to two typical axion models.
    Other lines show the upper limits and preliminary ones are in
    dashed lines.
    The new limit and the previous ones\cite{sumico1997,sumico2000}
    are plotted in blue lines.
    Limits by other axion helioscope experiments: Lazarus
    \etal~\cite{Lazarus} and CAST~\cite{CAST} are shown in red lines.
    Limits by axion Bragg scattering experiments \cite{Pascos}:
    SOLAX~\cite{solax1999}, COSME~\cite{cosme2002},
    DAMA~\cite{DAMA2001}, and CDMS~\cite{cdms2008} are shown in green
    lines.
    The black solid line shows the solar limit inferred from the solar
    age consideration.
    The right figure show the magnified view of the new limit.}
  \label{fig:exclusion}
\end{figure}

Fig.~\ref{fig:exclusion} shows the limit plotted as a function of
$m_a$.
The previous limits from the first \cite{sumico1997} and the second
\cite{sumico2000} phase measurements and some other bounds are
also plotted in the same figure.

\section{Conclusion}
The axion mass around 1\,eV has been scanned with an axion helioscope
with cold helium gas as the dispersion-matching medium in the
$4{\rm\,T}\times2.3\rm\,m$ magnetic field, but no evidence for solar
axions was seen.
A new limit on $\gagg$ shown in Fig. \ref{fig:exclusion} was set for
$0.84<m_a<1.00\rm\,eV$.
This is the first result to search for the axion in the $\gagg$-$m_a$
parameter region of the realistic axion models \cite{GUT_axion} with a
magnetic helioscope.  Full description of the present result is
published in Ref.~\cite{sumico2007}.

\acknowledgments The authors thank the former director general of KEK,
Professor H. Sugawara, for his support in the beginning of the
helioscope experiment.
This research was partially supported by the Japanese Ministry of
Education, Science, Sports and Culture, Grant-in-Aid for COE Research
and Grant-in-Aid for Scientic Research (B), and also by the
Matsuo Foundation.

\end{document}